\documentclass[%
 reprint,
superscriptaddress,
 amsmath,amssymb,
 aps,
]{revtex4-2}

\usepackage{graphicx}
\usepackage{dcolumn}
\usepackage{bm}
\usepackage[colorlinks = true, linkcolor = blue, urlcolor = blue, citecolor = blue, anchorcolor = blue]{hyperref}
\usepackage{xcolor}

\usepackage{newunicodechar}
\newunicodechar{σ}{$\sigma$}
\newunicodechar{θ}{$\theta$}
\newunicodechar{π}{$\pi$}

\newcommand{\blue}[1]{\textcolor{blue}{#1}}

\newcommand{\newtheta}{\vartheta}
\newcommand{\abs}[1]{\lvert #1 \rvert}
\newcommand{\kf}{k_{\mathrm{F}}}

\begin{document}

\preprint{APS/123-QED}

\title{Topologically enabled superconductivity}

\author{Michael A. Rampp}
\affiliation{%
Institute for Theory of Condensed Matter,
Karlsruhe Institute of  Technology, 76131 Karlsruhe, Germany
}%

\author{Elio J. K\"onig}
\affiliation{
 Max Planck Institute for Solid State Research, D-70569 Stuttgart, Germany
}%

\author{J\"org Schmalian}
\affiliation{%
 Institute for Theory of Condensed Matter,
Karlsruhe Institute of  Technology, 76131 Karlsruhe, Germany
}%
\affiliation{Institute for Quantum Materials and Technologies,
Karlsruhe Institute of  Technology, 76021 Karlsruhe, Germany}

\date{\today}
             
\begin{abstract}
Majorana zero modes are a much sought-after consequence of one-dimensional topological superconductivity. Here we show that, in turn, zero modes accompanying dynamical instanton events strongly enhance - in some cases even enable -   superconductivity. We find that the dynamics of a one-dimensional topological triplet superconductor is governed by a  $\theta$-term in the action. For isotropic triplets, this term enables algebraic charge-$2e$ superconductivity, which is destroyed by fluctuations in non-topological superconductors. For anisotropic triplets, zero modes suppress quantum phase slips and stabilize superconductivity over a large region of the phase diagram. We present  predictions of correlation functions and thermodynamics for states of topologically enhanced superconductivity.
\end{abstract}

\maketitle


One-dimensional topological $p$-wave superconductors are widely proposed 
as building blocks for quantum information processing~\cite{kitaev_unpaired_2001,Mourik2012,Sau2010,Oreg2010,Alicea2010,Beenakker2013,Albrecht2016,Sato2019}. While proximity-induced  superconductivity is discussed most frequently, the prospect of intrinsic superconductivity in one-dimensional structures would certainly allow for more versatile architectures. However, despite the obvious challenge of identifying the right material, there seems to be a more fundamental limitation to such an approach. In low dimensions the influence of fluctuations is very strong.  
The impact of order-parameter fluctuations on one-dimensional topological superconductivity has been extensively
studied for the spinless $p$-wave case~\cite{fidkowski2011majorana,sau2011number,kane_pairing_2017,keselman_one-dimensional_2018} and it has been demonstrated that vacuum tunneling by $2\pi$ quantum phase slips (QPS) is suppressed in these systems~\cite{sau2011number,pekker_suppression_2013}, thus enlarging the superconducting domain in the phase diagram. 
However, for isotropic triplet superconductivity the role of fluctuations in the spin sector is ordinarily so strong as to completely destroy $2e$ superconductivity~\cite{babaev_fractional-flux_2005,EsslerTsvelik2009,fradkin_colloquium_2015,fernandes_intertwined_2019}.

\begin{figure}[h!]
    \centering
   \includegraphics[width = .45\textwidth]{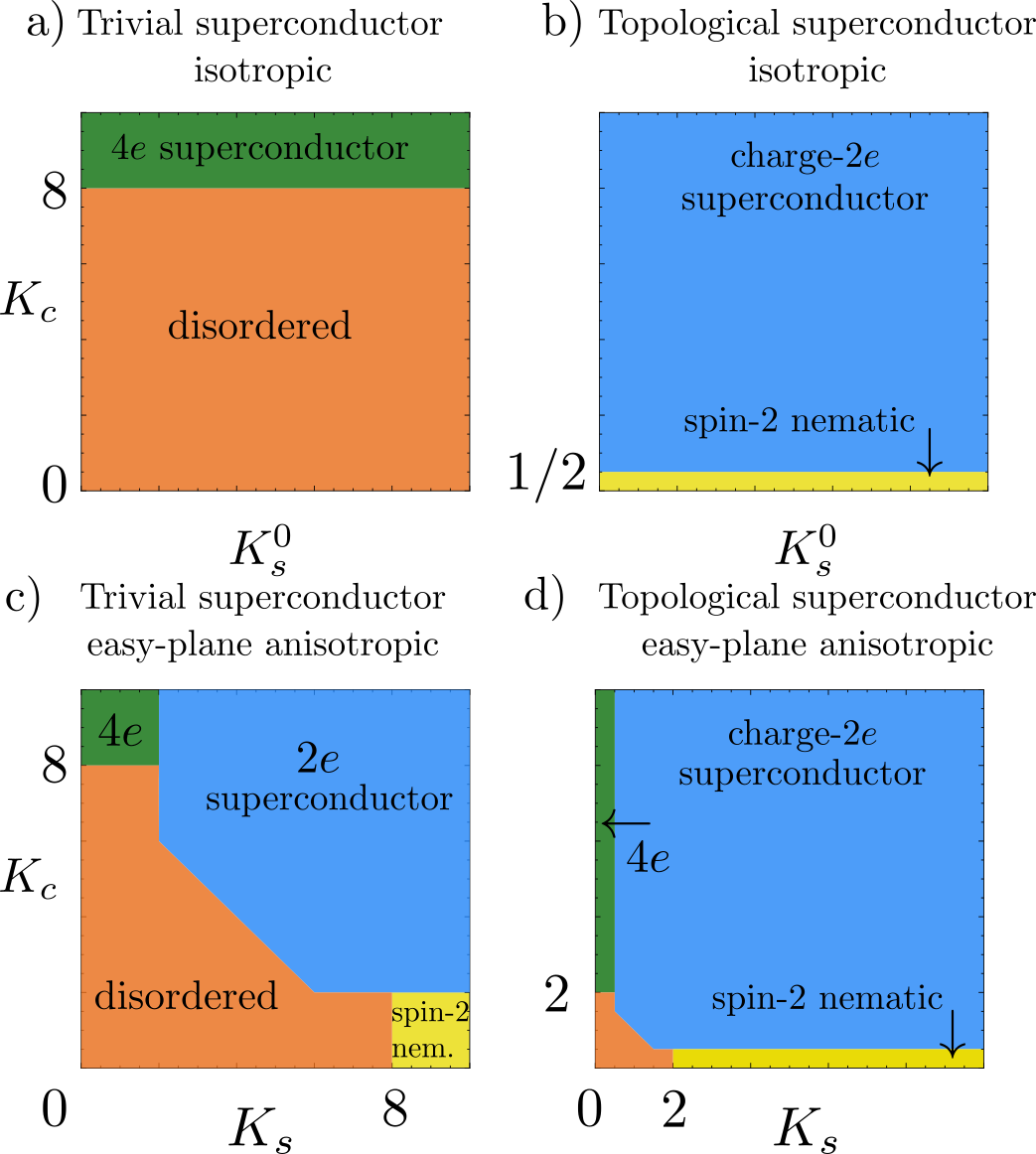}
   \caption{ Ground state phase diagram of the trivial (a), (c) and topological (b), (d) superconductor. The top (bottom) row depicts the spin-isotropic (strongly anisotropic) case. The axes correspond to the renormalized, i.e., experimentally relevant, charge ($K_c$) stiffness, and the bare ($K_s^0$) and renormalized ($K_s$) spin stiffness in the isotropic and the anisotropic case, respectively.
   All superconducting states have algebraic, quasi long-range order. While in the isotropic case  
   only vestigial charge-$4e$ superconductivity is allowed for topologically trivial systems, a $\theta$-term in the theory enables charge-$2e$ pairing of topological superconductors, which occurs in a much larger domain of the phase diagram. 
   At finite spin anisotropy, disorder-inducing quantum phase slips are suppressed by zero modes,  stabilizing superconductivity. 
   }
    \label{fig:Summary}
\end{figure}

In this paper we show that charge-$2e$ topological triplet superconductivity in one-dimensional quantum wires becomes  possible while it is not allowed for non-topological systems. The latter can only undergo vestigial charge-$4e$ pairing in a much reduced regime of the phase diagram, see Fig.~\ref{fig:Summary} a) and b). Zero modes, primarily discussed as static Majorana bound states  of topological superconductivity, emerge in our analysis as dynamical events accompanying instantons of the order parameter field. 
They are shown to  suppress order-parameter fluctuations via destructive interference due to a Berry phase, enabling the charge-$2e$ superconducting state.
This Berry phase leads to a  topological term in the field theory, a $\theta$-term~\cite{abanov_theta-terms_2000,altland_condensed_2010}. For the topological angle we find $\theta=\pi$ for topological superconductors and $\theta=0$ for non-topological ones. We obtain this result using non-Abelian bosonization and, using a physically more transparent reasoning, by demonstrating that dynamical  zero modes give rise to a  complex 
QPS fugacity.

These findings occur for systems that are isotropic in spin space where we exploit a connection to Haldane's conjecture~\cite{haldane_nonlinear_1983} for spin chains. 
Including an anisotropy in spin space, disorder-inducing configurations are  suppressed by zero modes that are more akin to what is known for spinless systems~\cite{sau2011number,pekker_suppression_2013}.

\emph{\blue{Model:}} We study a time reversal invariant $p$-wave spin triplet superconductor in one dimension at zero temperature. The Bogoliubov-de Gennes Hamiltonian for given space- and time-dependent order-parameter configuration $\boldsymbol{\Delta}\left(x,\tau\right)$ in the Nambu spinor basis $\Psi=(\psi_{\uparrow},\psi_{\downarrow},\bar \psi_{\downarrow},-\bar \psi_{\uparrow})^{T}$ is
\begin{equation}
{\cal H}\left[\boldsymbol{\Delta}\right]=\left(\begin{array}{cc}
-\frac{\partial_{x}^{2}}{2m}-\mu & \boldsymbol{\Delta}\left(x,\tau\right)\cdot\boldsymbol{\sigma} \partial_{x}\\
-\partial_{x}\boldsymbol{\Delta}^{\dagger}\left(x,\tau\right)\cdot\boldsymbol{\sigma} & \frac{\partial_{x}^{2}}{2m}+\mu
\end{array}\right),
\label{BdG}
\end{equation}
with $m$ the fermion mass and $\mu$ the chemical potential.
The order parameter takes the form of a real unit vector $\bm{n}$, that describes the orientation of the Cooper pair spin, times a global phase $\newtheta$, \emph{i.e.}  $\bm{\Delta}=\abs{\Delta} e^{i\newtheta}\bm{n}$, with the pairing strength $\abs{\Delta}$. 
$\mu<0$ corresponds to  a trivial and  
$\mu>0$ to a topological superconductor~\cite{ReadGreen}.
The order parameter manifold is  $\mathcal M = \left(S^{1}\times S^{2}\right)/\mathbb{Z}_{2}$, where the $\mathbb Z_2$ quotient stems from the equivalence of field configurations $(\newtheta, \bm{n}) \sim (\newtheta + \pi, - \bm{n})$~\cite{korshunov_two-dimensional_1985,mukerjee_topological_2006}.
The fermionic dynamics is then governed by the action 
\begin{equation}
S_{f}=\frac{1}{2}\int \mathrm{d}\tau \mathrm{d}x \Psi^{\dagger}\left({\cal \partial_{\tau}+{\cal H}}\right)\Psi.    
\label{eq:fermions}
\end{equation}
In low-dimensional intrinsic superconductors order-parameter fluctuations are important and are governed by the non-linear $\sigma$-model (NL$\sigma$M):
\begin{eqnarray}
	S_b &=& \frac{K_{c}^{0}}{2\pi}  \int \mathrm{d}\tau \mathrm{d}x \left( \frac{1}{v_{c}}\left(\partial_{\tau}\newtheta\right)^{2}+ v_{c}\left(\partial_{x}\newtheta\right)^{2} \right) \nonumber\\
	& +& \frac{K_{s}^{0}}{2\pi} \int \mathrm{d}\tau \mathrm{d}x \left( \frac{1}{v_{s}}\left(\partial_{\tau}\bm{n}\right)^{2} + v_{s}\left(\partial_{x}\bm{n}\right)^{2} \right) . \label{eq:bosons}
\end{eqnarray}
Here $K_{c}^{0}$ and $K_{s}^{0}$ denote the bare stiffnesses of the charge and spin sector respectively. 
$v_{c}$ and $v_{s}$ are the respective velocities. We use $K_{c,s}$ for the renormalized stiffnesses.

This model is a generalization  of previous descriptions for spinless fermions~\cite{MeyerLarkin2007,SitteGarst2009,kane_pairing_2017}.

\begin{figure*}
    \centering
    \includegraphics[width = .95\textwidth]{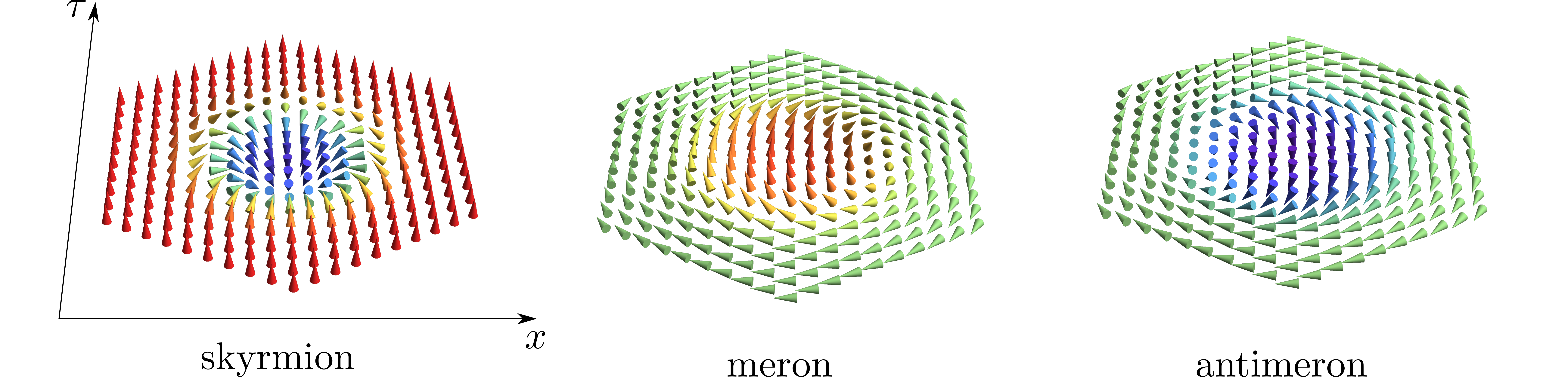}
    \caption{Selected topologically nontrivial order-parameter configurations in space-time~\cite{link_mathematica}.
    Left panel:
    skyrmion of topological charge $Q=1$.  Middle and right panel $Q=\pm \tfrac{1}{2}$ merons, i.e. spin vortices that differ by the orientation of the spin inside the core.  }
    \label{fig:Topological defects}
\end{figure*}

\emph{\blue{Topological field configurations:}} 
The fields $\newtheta, \bm{n}$ that describe $U(1)$-phase and spin  of  a Cooper pair allow for several distinct QPSs, shown in Fig.~\ref{fig:Topological defects}: we call vortices of $\newtheta$ with winding $\nu_c$ in space-time  $2\pi \nu_c$ \textit{charge QPSs}. Space-time \textit{skyrmions} in $\bm{n}$ have integer winding $Q$ which relies on $\pi_2(S^2) = \mathbb Z$. 
The evaluation of the partition function contains a sum over all possible topological sectors, determined by a set of numbers $N=\{\nu_c,Q,\cdots\}$~\cite{altland_condensed_2010}
\begin{equation}
    Z = \sum_{N} \int \mathcal D\newtheta_N \mathcal D \bm{n}_N Z_N^f[\newtheta, \bm{n}] e^{- S_b[\newtheta, \bm{n}]} \label{eq:SumTopoSectors},
\end{equation}
where $\int D\newtheta_N \mathcal D \bm{n}_N \dots$ denotes the integral over smooth bosonic fluctuations on top of the topological field configuration and $Z_N^f[\newtheta, \bm{n}]$ is the fermionic partition sum in a given bosonic background of fixed $N$. Notice, all fluctuations around the trivial state, including these instantons, must be taken into account as long as their action is finite.
Estimating the core size of QPS we obtain, following Ref.~\cite{zaikin1997quantum}, $r_{\rm QPS}\approx\left(K_c v/2\pi E_{\mathrm{cond}}\right)^{1/2}$, and the core action $S_{\rm QPS}\approx K_c/2$.

\emph{\blue{Topologically trivial superconductivity:}} To analyze the model it is tempting to argue that fermions are gapped and should not change the universal behavior of order-parameter fluctuations. Then, 
the NL$\sigma$M of Eq.~\eqref{eq:bosons} implies $\langle \bm{\Delta} \rangle=0 $ with exponentially decaying correlations, caused by fluctuations in the spin sector. As last resort the system can still enter a state of algebraic vestigial order characterized by the composite $\bm{\Delta} \cdot\bm{\Delta} $ below a Berezinskii-Kosterlitz-Thouless (BKT) transition~\cite{korshunov_two-dimensional_1985,mukerjee_topological_2006}. However, this state, where two spin triplets form a charge-$4e$ spin singlet, can be destroyed by the proliferation of $\pi$, i.e. fractional QPSs. In our units, the QPS configurations with smallest winding number $\nu_c$ become relevant at a critical stiffness of $K_c=2/\nu_c^2$. Hence, in order to stabilize $4e$ order with $\nu_c=1/2$ fractional vortices, the charge stiffness has to be four times larger compared to the usual BKT transition, which implies $K_c=8$; see the left panel of Fig.~\ref{fig:Summary}. 
This behavior, deduced from the two-dimensional classical model, is indeed correct in the topologically trivial phase. Yet, as we will see next, it does not apply to topological superconductors. 
In the topological phase order-parameter configurations with nontrivial topology in space-time play a key role\cite{erten_skyrme_2017,chatterjee_superconductivity_2016,chatterjee_intertwining_2017,choi_topological_2018,senthil_symmetry-protected_2015,turner_topological_2011,sedrakyan_topological_2017}, requiring us to carefully distinguish their effects and their interplay with the fermionic degrees of freedom.

\emph{\blue{Integrating out fermions:}} 
In order to systematically integrate out fermions in a one-dimensional system we employ non-Abelian bosonization~\cite{witten_non-abelian_1984}, that manifestly preserves the symmetry properties of interacting fermion theories and is expressed in  terms of the 
Wess-Zumino-Novikov-Witten (WZNW) action. We employ this method to effectively eliminate the massive degrees of freedom while retaining the effect of zero modes. The degrees of freedom of the WZNW theory are group valued boson fields defined on the orthogonal group $O(4)$. The matrix components of this field can be related to fermionic currents by $g_{ij}=(-i/M)\chi_{\rm{R}}^{i}\chi_{\rm{L}}^{j}$, where $\chi_{\rm{R}},\chi_{\rm{L}}$ are right- and left-moving Majorana fermions that can be constructed from the microscopic fermions, and $M$ is a regularization dependent mass-scale. The degrees of freedom described by the BdG-Hamiltonian~\eqref{BdG} deep in the topological phase are exactly four right-moving and four left-moving Majorana fermions, which explains the manifold $O(4)$. Effectively, the field $g$ encodes the fermionic fluctuations and the opening of a gap corresponds to breaking the chiral symmetry, \emph{i.e.} individual components of $g$ develop a finite expectation value. The action with coupling constant $\gamma$ reads
\begin{equation}
    S_W=\frac{1}{4 \gamma^2}\int \mathrm{d}\tau \mathrm{d}x  {\rm tr}\left[\partial_\mu g \partial^\mu g^{-1}\right]+\Gamma,
    \label{eq:WZNW}
\end{equation}
where the Wess-Zumino term $\Gamma [g]$ is defined by extending the domain of the group valued field $g\in O(4)$ to a hemisphere of the three-dimensional unit sphere with $S^2$ as boundary
\begin{equation}
    \Gamma=\frac{\varepsilon^{\mu \nu \rho}}{24\pi i}
    \int
    \mathrm{d}^3r
 \operatorname{ tr}\left[g^{-1}(\partial_\mu g) g^{-1}(\partial_\nu g) g^{-1}(\partial_\rho g) \right].
\label{eq:WZ}
\end{equation}
 
The pairing term is bosonized and introduces a $(\newtheta,\bm{n})$ dependent mass term for $g$. In the adiabatic limit $g$ follows the variation of the background fields $(\newtheta,\bm{n})$ and thereby traces the equator of $O(4)$. The details of the subsequent analysis are given in the supplement~\cite{SuppMat}. The first term in \eqref{eq:WZNW} yields terms identical to the NL$\sigma$M in  \eqref{eq:bosons}, i.e. it yields a renormalization of the stiffnesses due to fermions. More interesting is the second term. When constrained to the equator, the Wess-Zumino term \eqref{eq:WZ}, which measures the solid hyperangle on $O(4)$ normalized to $2\pi$, may only take values of $\pi \,\text{mod}\, 2\pi$. We find  that 
\begin{equation}
\Gamma \rightarrow S_{\theta} = i \frac{\theta}{4\pi} \int \mathrm{d}\tau \mathrm{d}x \bm n \cdot (\partial_\tau \bm n \times \partial_x \bm n) ,
\label{eq:Stheta_from_WZ}
\end{equation}
where $\theta=\pi$. For a topologically trivial superconductor we obtain instead $\theta=0$.
 Eq.~\eqref{eq:Stheta_from_WZ} has profound implications. According to Haldane's conjecture~\cite{haldane_nonlinear_1983} where $\theta=\pi$ occurs in the same action for $\bm n $   for half-integer spins~\cite{fradkin_topological_1988}, $S_\theta$ leads to a critical state described by a $SU_1(2)$ WZNW theory~\cite{affleck_exact_1986,affleck_critical_1987}, rather than a state with finite correlation length. Hence, algebraic charge-$2e$ superconductivity becomes possible for topological superconductors while it is forbidden in topologically trivial ones.
 
 Before we discuss further implications of this finding, we offer an alternative and physically more transparent derivation of the $\theta$-term.
 We consider nontrivial skyrmion configurations but 
in this context it is particularly convenient to introduce
 a soft easy-plane anisotropy~\cite{affleck_mass_1986}. This anisotropy yields spin-vortex configurations  but allows $\bm{n}$ to escape the plane in a core region, whose size is determined by the strength of the anisotropy, avoiding a singularity. There are two kinds of $2\pi$ spin QPSs distinguished by the orientation of the vector in the core $(\pm  n_z)$,  called \textit{meron} and \textit{antimeron} with skyrmion winding number $Q=\pm \tfrac{1}{2}$; see Fig.~\ref{fig:Topological defects}.
 The fermionic operator ${\cal K}\equiv \partial_{\tau}+{\cal H}$ acts in space and (imaginary) time. Since the Nambu spinor satisfies the reality condition $\Psi^{T}=\Psi^{\dagger}C$ (where  $C=\sigma^{y}\tau^{y}$ with $\sigma^ {i}$ acting on spin and $\tau^{i}$ on Nambu degrees of freedom), the fermionic  partition function $Z_N^f[\newtheta, \bm{n}] = \operatorname{Pf}(C\mathcal{K}[\newtheta, \bm{n}])$  follows directly from the Pfaffian of the fermionic kernel. Hence, in the case that the fermionic kernel possesses a zero eigenvalue mode the fermionic partition function vanishes.
As summarized in~\cite{SuppMat}, integrating out fermions yields in the limit of large anisotropy 
that two zero modes are shifted to finite eigenvalue in such a manner that
\begin{equation}
Z_N^f[\newtheta, \bm{n}]\propto e^{-i \pi Q}. 
\label{eq:meron}
\end{equation}
With this complex fugacity, the contribution of $2\pi$ spin QPS vanishes upon summation over $Q$, the internal degree of freedom of the vortex~\cite{affleck_mass_1986}. Within a field theoretical language, this behavior is precisely the effect that follows from a $\theta$-term in the action, $S_\theta=i\theta Q$, where $ Q =\int \mathrm{d}\tau \mathrm{d}x \bm n \cdot (\partial_\tau \bm n \times \partial_x \bm n)/(4\pi)$ is the associated topological charge~\cite{altland_condensed_2010}. It reveals  that the $\theta$-term yields destructive interference of disordering spin configurations.

In addition to nontrivial spin textures, we can also analyze charge vortices or combinations of charge and spin vortices. We find that one can map this dynamic problem onto an effective Hermitian single-particle Hamiltonian $H_{\rm eff}$ in two dimensions $(x,\tau)$. 
$H_{\rm eff}$ can then be reduced to the two-dimensional Fu-Kane Hamiltonian of a 3D topological insulator surface state  in contact with an s-wave superconductor~\cite{fu_superconducting_2008}. 
Established results for zero modes due to static vortex configurations of the Fu-Kane Hamiltonian 
 can now be used to obtain the number and character of zero modes for ${\cal K}$.
Since one coordinate of this two-dimensional problem corresponds to (Euclidean) time,  those are again dynamical instanton events.   The details of this rather powerful but straightforward analogy are summarized in the supplementary material~\cite{SuppMat}.
It yields for example two zero modes of ${\cal K}$ for a charge $2\pi$ QPS ($\nu_c=1$) and one zero mode
for a combined $2\pi$  half QPS ($\nu_c=\nu_s=\tfrac{1}{2}$), with winding number of a planar spin vortex $\nu_s$.  More generally we obtain a zero mode for each odd $\nu_c\pm\nu_s$.  
These dynamical zero modes can be understood as protected level crossings under the adiabatic variation of a parameter. Perturbations that do not destroy the level crossing also do not lift the dynamic zero modes. 
Hence $Z_N^f[\newtheta, \bm{n}] =0$ for these single-defect configurations and the corresponding vortex fugacities  vanish. More importantly, vortex-antivortex pairs do not contribute to a BKT  transition as the exponentially small overlap between  modes gives rise to a linear, confining potential 
overruling the usual logarithmic interaction. Our mapping to the Fu-Kane model can also be applied to spinless $p$-wave superconductors
where it agrees with past results on  QPSs in this system~\cite{sau2011number,pekker_suppression_2013}.

\emph{\blue{Quantum Phase diagram.}}
If we combine the $\theta$-term and the presence of zero modes due to charge vortices, we can  determine the phase diagrams shown in Fig.~\ref{fig:Summary}. 
In the topologically trivial phase $2e$ superconducting order is destroyed because $\bm{n}$ fields are gapped but $4e$ superconducting order may exist for $K_c >8$. In contrast, in the topological phase the $\bm{n}$-fields are algebraically ordered and the charge-$2e$ superconductor persists. A BKT transition driven by $\nu_c=2$, i.e. $4\pi$, QPS disorders the phase sector for $K_c<2/\nu_c^2=1/2$.  
The resulting state possesses vestigial order of the composites $\Delta_{x}^{\dagger}\Delta_{y}+\Delta_{y}^{\dagger}\Delta_{x}$, $\Delta_{x}^{\dagger}\Delta_{x}-\Delta_{y}^{\dagger}\Delta_{y}$, and $\Delta_z^\dagger\Delta_z$. It is a charge insulator and a spin-nematic state. 
Hence, the critical stiffness for the destruction of superconductivity is $16$ times lower than 
in the trivial case; see Fig.~\ref{fig:Summary} b.

\emph{\blue{Correlators:}} Let us discuss experimentally and numerically measurable consequences of the topological phases in terms of bosonic and fermionic correlators. 
In topologically trivial superconductors, fermions are gapped and  fermion correlators decay exponentially. 
This is qualitatively different for topological superconductors. The power-law behavior of the single-fermion correlator is again caused by dynamical zero modes. We obtain
\begin{equation}
\langle\psi_{\sigma}(r)\psi_{\sigma'}^{\dagger}(0)\rangle \sim \delta_{\sigma\sigma'}{r}^{-\frac{1}{8K_{c}}-\frac{K_{c}}{2}-\frac{1}{2}},
\label{eq:fermion_powerlaw}
\end{equation}
where ${r}=\sqrt{x^2+v_c^2\tau^2}$ denotes the Euclidean norm of a point $(x,\tau)$ in space-time. For simplicity we assumed $v_s=v_c$.
Hence, the topological superconductor possesses gapless charged fermionic excitations. This is in contrast to nodal excitations which can appear in higher dimensions and are charge neutral. However, experiments which measure the phase coherence, such as flux quantization or the a.c. Josephson effect are expected to detect a condensate charge of $2e$.

Eq.~\eqref{eq:fermion_powerlaw} implies a power-law dependence on energy in the tunneling density of states $\rho(E)$, which can be detected in scanning tunneling microscopy measurements,
\begin{equation}
    \rho(E) \sim E^{\frac{1}{8K_{c}}+\frac{K_{c}}{2}-\frac{1}{2}}, \quad E>0.
\end{equation}
Similarly, we expect power-law dependence on temperature of various transport coefficients, such as the thermal conductivity~\cite{kane1996thermal}.

Notice that nonetheless the system cannot be described as a Tomonaga-Luttinger liquid of charge-$e$ fermions, since the behavior of two-particle correlators differs. 
Zero modes also nontrivially affect two-particle correlators in the topological state~\cite{SuppMat}.
Specifically, we find the tunneling density of states for tunneling a singlet and triplet Cooper pair into the wire 
\begin{eqnarray}
 \rho_{\rm{sg}}(E) &\sim& E^{2K_c+\frac{1}{2K_{c}}-1}, \nonumber \\
 \rho_{\rm{tr}}(E) &\sim& E^{\frac{1}{2K_{c}}}.
\end{eqnarray}
The exponent for the singlet pair-correlator differs from the Luttinger liquid result (extracting the Luttinger parameters from the one-particle correlator) for general values of $K_c$, $\rho_{\rm{sg}}^{\rm{LL}} \sim E^{1+\frac{1}{2K_{c}}}$. In the $2e$-ordered phase the singlet correlations are less singular than in a conventional Luttinger liquid. At low energies the amplitude for tunneling triplet pairs into the system is enhanced compared to the one for singlet pairs.
A  detailed derivation is given in~\cite{SuppMat}.

\emph{\blue{Easy axis anisotropy:}} We already discussed an additional easy axis anisotropy $S_{\mathrm{an}} = \lambda \int \mathrm{d}\tau \mathrm{d}x  n_{z}^{2}$ when we offered an alternative derivation of the $\theta$-term in Eq.~\eqref{eq:meron}. The fact that topological superconductivity is stabilized over the topologically trivial case can also be seen at $\lambda>0$. It is however more similar to what is known from spinless $p$-wave superconductors~\cite{sau2011number,pekker_suppression_2013},  since the spin sector is now also governed by a $U(1)$ order parameter and charge-$2e$ superconductivity becomes possible for both topological and non-topological superconductors. 
However, at small stiffnesses  algebraic order is now destroyed by the proliferation of three kinds of space-time topological defects:  $2\pi$ charge, spin, or combined vortices; see also Refs.~\cite{KruegerScheidl2002,PodolskyVishwanath2009,KoenigPixley,chung2021berezinskii}.
This yields the four possible phases presented in Fig.~\ref{fig:Summary} (c and d): a completely ordered (all QPSs expelled), a completely disordered (all QPSs proliferate), and two vestigial phases (only one kind of QPS proliferates).  Physically the resulting phases can be identified as: a spin nematic charge-$2e$ superconductor, where the two gapless excitations are spin-$1$ Cooper pairs, a correlated insulator, a charge-$4e$ superconductor with gapless excitations carrying charge $4e$ and spin $0$, and a spin-nematic phase similar to the isotropic case. 
While thermodynamically the phase diagram of topological and trivial limits are similar,  in the former case  $2\pi$ charge, spin, or combined QPSs are suppressed by dynamic zero modes. Hence, the leading transitions are effected only by $4\pi$ QPSs, and as a result we find again  that the superconducting state is stabilized for the topological phase. Similar to the isotropic case, zero modes lead to gapless fermion and two-particle excitations in the $2e$ phase. In the vestigial charge-$4e$ phase, single fermions are gapped, but the singlet pair-correlator remains gapless. There is another distinction between the topological and the trivial vestigial phases: in the topological case the bosonic operator $\phi_s(x)$ ($\phi(x)$), which is the dual field to the phase $\newtheta_s(x)$ ($\newtheta(x)$), develops \textit{bona-fide} long range $\mathbb Z_2$ order in the vestigial charge (spin) phase on the top left (bottom right) of the phase diagram. The reason is that the partial disorder is induced by $4\pi$ spin (charge) QPSs that preserve a remnant $\mathbb{Z}_{2}$ order. Physically, these order parameters correspond to the magnetization (charge) integrated up to a point $x$ along the 1D system - thus they are non-local order parameters.

\emph{\blue{Conclusion:}} We find that superconducting fluctuations in topological and topologically trivial superconductors are qualitatively different, leading to distinct phase boundaries, symmetry breaking, and excitation spectra. The common theme is that superconductivity in topological systems is much more robust against fluctuations. The reason is the crucial role of dynamic zero modes. The most dramatic effect is the emergence of otherwise forbidden charge $2e$ superconductivity for isotropic triplets caused by a topological term in the action. It yields algebraic superconducting order for a stiffness 16 times lower than charge $4e$-superconductivity in the non-topological counterpart.

\emph{\blue{Acknowledgements:}} We are grateful to P. Coleman, R. M. Fernandes, L. I. Glazman, P. Goswami, Y. Komijani, Y. Oreg, A. Schnyder, and A. Shnirman for useful discussions.
M. A. R. and J. S. acknowledge support by the Deutsche Forschungsgemeinschaft (German Research Foundation) Project ER 463/14-1.  The work by E. J. K. was partly performed at the Aspen Center for Physics, which is supported by National Science Foundation grant PHY-1607611.

\bibliography{bibliography_top_enh}


\setcounter{equation}{0}

\setcounter{figure}{0}

\setcounter{table}{0}

\setcounter{section}{0}

\makeatletter

\renewcommand{\theequation}{S\arabic{equation}}

\renewcommand{\thefigure}{S\arabic{figure}}

\renewcommand{\thesection}{S-\Roman{section}}

\renewcommand{\bibnumfmt}[1]{[S#1]}

\renewcommand{\citenumfont}[1]{S#1} 


\begin{widetext}

\setcounter{page}{1}

\begin{center}

Supplementary materials on \\

\textbf{Topologically enabled superconductivity}\\

Michael A. Rampp$^{1}$, Elio J. K\"onig$^{2}$, Jörg Schmalian$^{1,3}$\\ 

$^{1}${Institut für Theorie der Kondensierten Materie,
Karlsruher Institut für Technologie, 76131 Karlsruhe, Germany}\\

$^{2}${Max Planck Institute for Solid State Research, D-70569 Stuttgart, Germany}\\

$^{3}${Institut für Quantenmaterialien und Technologien,
Karlsruher Institut für Technologie, 76021 Karlsruhe, Germany}

\end{center}

\end{widetext}

\section{Computation of zero modes}

\subsection{Construction of the $(2+0)$-dimensional effective Hamiltonian}

In the following it will be argued that the computation of the Pfaffian of the fermionic kernel $\mathcal{K}$ (to be defined below) of a dynamical problem in $(1+1)$-dimensions can be mapped to a time-independent Hamiltonian problem in two spatial dimensions. Subsequently, the zero modes of this Hamiltonian will be found for the topological background configurations introduced in the main text. The kernel $\mathcal{K}$ is not Hermitian. This prohibits in general an expansion in a complete set of eigenmodes. By multiplying $\mathcal{K}$ with a matrix such that the resulting operator is Hermitian, this issue can be circumvented. This procedure allows to determine the value of the Pfaffian up to a phase, since $\operatorname{Pf}(AB)^{2}=\det(A)\det(B)$, which follows from $\operatorname{Pf}(A)^{2}=\det(A)$~\cite{shankar_quantum_2017} and multiplicativity of the determinant.

Let us first construct the kernel $\mathcal{K}$. Consider the fermions to be deep in the topological superconducting phase $\mu\gg2m\abs{\Delta}^{2}$. Then the low-energy degrees of freedom are situated near the Fermi points. Define the real and imaginary parts of the components of the gap function as $\Delta'_{j} \equiv \operatorname{Re}\abs{\Delta}k_{\mathrm{F}}e^{i\newtheta}n_{j}$ and $\Delta''_{j} \equiv \operatorname{Im}\abs{\Delta}k_{\mathrm{F}}e^{i\newtheta}n_{j}$. We expand close to the Fermi points $\pm k_{\mathrm{F}}=\pm\sqrt{2m\mu} = \pm m v_F$. The Nambu spinors now have eight components corresponding to spin, Nambu, and right/left moving components, which we describe using the Pauli matrices $\sigma^i$, $\tau^i$, and  $\rho^i$, respectively. The reality condition then takes the form $\Psi^{\dagger}=\Psi^{T}\sigma^{y}\tau^{y}\rho^{x}\equiv\Psi^{T} C$. (The charge-conjugation operator $C$ also exchanges left- and right-movers). Then for $\mathcal{K}\equiv (\partial_{\tau}+\mathcal{H})$, holds $Z^f[\newtheta, \bm{n}] = \rm{Pf}(C\mathcal{K}[\newtheta, \bm{n}])$. Explicitly we have
\begin{eqnarray}
 C \mathcal K &=& \sigma^{y}\tau^{y}\rho^{x}( \partial_{\tau} + v_{\mathrm{F}}(-i\partial_{x}) \tau^{z}\rho^{z} \nonumber\\& &+ \Delta'_{j}\sigma^{j}\tau^{y}\rho^{z} + \Delta''_{j}\sigma^{j}\tau^{x}\rho^{z}). \label{eq:kernel_lin}
\end{eqnarray}
Note that $C$ and $\mathcal K$ have a different matrix structure than in the main text, since the expansion around the Fermi points was not explicitly performed there. The sign of the gap function at the left Fermi point is reversed, due to the antisymmetry of the $p$-wave pairing in momentum space. The term $\sim\partial_{x}(u e^{-i\newtheta}{\bm{n}})$ has been neglected, because typical bosonic momenta are much smaller than $\kf$.

In order to map this problem to an Hermitian operator, we multiply by the matrix $\sigma^{y}\tau^{y}\rho^{z}$ and obtain
\begin{eqnarray}
H_{\mathrm{eff}}\equiv \sigma^{y}\tau^{y}\rho^{z} C\mathcal K &=& i\rho^{y}\partial_{\tau} + i\tau^{z}\rho^{x}v_{\mathrm{F}}\partial_{x} \nonumber\\& &- \Delta'_{j}\sigma^{j}\tau^{y}\rho^{x} -\Delta''_{j}\sigma^{j}\tau^{x}\rho^{x}. \label{eq:effective}
\end{eqnarray}
This is an effective Hamiltonian of a $(2+0)$-dimensional system with coordinates $(x,\tau)$. It has a chiral symmetry $\mathcal{C}=\rho^{z}$. The kinetic term has the structure of a Dirac Hamiltonian with $\alpha$-matrices $\alpha^{x}=-\tau^{z}\rho^{x}$ and $\alpha^{\tau}=-\rho^{y}$ that satisfy the Clifford algebra $\{\alpha^{i},\alpha^{j}\}=2\delta_{ij}$. It takes the form
\begin{equation}
    H_{\mathrm{eff,kin}} = -iv_{\mathrm{F}}\alpha^{x}\partial_{x} -i\alpha^{\tau}\partial_{\tau}.
\end{equation}
The components of the gap function act as coordinate dependent mass terms. It is convenient to introduce polar coordinates in the $(x,\tau)$-plane by $x=r\cos\phi,\,v_{\mathrm{F}}\tau=r\sin\phi$, and associated Dirac $\alpha$-matrices $\alpha^{r}=\cos\phi\alpha^{x}+\sin\phi\alpha^{\tau},\,\alpha^{\phi}=-\sin\phi\alpha^{x}+\cos\phi\alpha^{\tau}$. In these coordinates the effective Hamiltonian takes the form
\begin{eqnarray}
H_{\mathrm{eff}} &=& -iv_{\mathrm{F}}\alpha^{r}\partial_{r} -i\frac{\alpha^{\phi}}{r}\partial_{\phi}- \Delta'_{j}(r,\phi)\sigma^{j}\tau^{y}\rho^{x} \nonumber\\&-&\Delta''_{j}(r,\phi)\sigma^{j}\tau^{x}\rho^{x}. \label{eq:effective_polar}
\end{eqnarray}

\subsection{Zero Modes on QPSs}

Consider now a charge QPS with winding number $\nu_c\in\mathbb{Z}$ situated at the origin. This corresponds to a configuration of the gap function where $\Delta_{j}'=\Delta_{0}(r)\cos(\nu_c\phi)n_{j}$ and $\Delta_{j}''=\Delta_{0}(r)\sin(\nu_c\phi)n_{j}$.  The real and positive amplitude of the $\Delta_{0}(r)$ has to vanish at the origin for the gap function to be well-defined. Far away from the core the amplitude approaches the equilibrium value $\abs{\Delta}k_{\mathrm{F}}$. The $\bm{n}$-vector can be chosen to point in any constant direction, and we choose the $z$-direction $\bm{n}=\bm{e}_{z}$. Hence, for this particular configuration the effective Hamiltonian reads
\begin{eqnarray}
H_{\mathrm{eff}} &=& -iv_{\mathrm{F}}\alpha^{r}\partial_{r} -i\frac{\alpha^{\phi}}{r}\partial_{\phi}- \Delta_{0}(r)\cos(\nu_c\phi)\sigma^{z}\tau^{y}\rho^{x} \nonumber\\&-&\Delta_{0}(r)\sin(\nu_c\phi)\sigma^{z}\tau^{x}\rho^{x}. \label{eq:FK_charge_QPS}
\end{eqnarray}
This Hamiltonian is identical to a direct sum of two (2+0)-dimensional Fu-Kane Hamiltonians of a 3D topological insulator surface state in contact with an s-wave superconductor~\cite{fu_superconducting_2008}. The decomposition is with respect to the two spin-sectors that are defined by the choice of the constant vector $\bm{n}$ which defines a quantization axis. It is known that the Fu-Kane Hamiltonian possesses a single localized zero mode in a background field with odd winding number and no zero mode in a background field with even winding number. We can conclude that the Hamiltonian~\eqref{eq:FK_charge_QPS} possesses two localized zero modes in the field of a charge QPS of odd winding number. Moreover, there exists an anti-unitary operator that commutes with~\eqref{eq:FK_charge_QPS} and squares to $-1$, an effective time-reversal symmetry, given by $\mathcal{T}=\sigma^{x}\tau^{y}\mathfrak{K}$. This enforces the double-degeneracy of the spectrum, while the particle-hole symmetry $\mathcal{P}$ defined by $\mathcal{C}=\mathcal{PT}$ protects the zero modes.

The solutions can also be found explicitly for $\nu=\pm 1$ by looking for angle-independent solutions to the zero mode equation. For $\nu_c=+1$ we find, introducing $g_{c}(r)= N\exp\left(-\int_{0}^{r}\mathrm{d}r'\Delta_{0}(r')/v_{\mathrm{F}}\right)$,
\begin{subequations}
\begin{eqnarray}
\Psi_{1}^{T}(r) &=& g_{c}(r)\frac{1}{\sqrt{2}}\begin{pmatrix}1&0& 1&0& 0&0& 0& 0\end{pmatrix}, \\
\Psi_{2}^{T}(r) &=& g_{c}(r)\frac{1}{\sqrt{2}}\begin{pmatrix}0&1& 0&1& 0&0& 0& 0\end{pmatrix}.
\end{eqnarray}
\end{subequations}

For the case of $\nu_c=-1$ we find analogously
\begin{subequations}
\begin{eqnarray}
\Psi_{1}^{T}(r) &=& g_{c}(r)\frac{1}{\sqrt{2}}\begin{pmatrix}0&0& 0&0& 1&0& 1& 0\end{pmatrix}, \\
\Psi_{2}^{T}(r) &=& g_{c}(r)\frac{1}{\sqrt{2}}\begin{pmatrix}0&0& 0&0& 0&1& 0& 1\end{pmatrix}. 
\end{eqnarray}
\end{subequations}

One easily realizes that the conclusions made about the zero modes do not rely on our simplifying assumption that $v_\mathrm{F}=v_c$.

\subsection{Zero mode on combined QPSs}

Consider now a combined charge and spin QPS, that is a configuration in which both the phase and the $\bm{n}$-vector wind around the origin with winding numbers $\nu_{c}$ and $\nu_{s}$ respectively. Note that if both $\nu_{c}$ and $\nu_{s}$ are half-integer the gap function is still well-defined. We choose the $\bm{n}$-vector to lie in the $xy$-plane, thus the components of the gap function take the form $\Delta_{x}'=\Delta_{0}(r)\cos(\nu_{c}\phi)\cos(\nu_{s}\phi)$, $\Delta_{y}'=\Delta_{0}(r)\cos(\nu_{c}\phi)\sin(\nu_{s}\phi)$, $\Delta_{x}''=\Delta_{0}(r)\sin(\nu_{c}\phi)\cos(\nu_{s}\phi)$, and $\Delta_{y}''=\Delta_{0}(r)\sin(\nu_{c}\phi)\sin(\nu_{s}\phi)$. The remaining components vanish. The resulting effective Hamiltonian reads
\begin{widetext}
\begin{eqnarray}
H_{\mathrm{eff}} = -iv_{\mathrm{F}}\alpha^{r}\partial_{r} -i\frac{\alpha^{\phi}}{r}\partial_{\phi}  - \Delta_{0}(r)\begin{pmatrix} 0 & e^{-i\nu_{s}\phi}\\ e^{i\nu_{s}\phi}& 0\end{pmatrix}\otimes\begin{pmatrix} 0 & -ie^{i\nu_{c}\phi}\\ ie^{-i\nu_{c}\phi}& 0\end{pmatrix}\otimes\rho^{x} . \label{eq:FK_combined_QPS}
\end{eqnarray}
\end{widetext}
It can be seen that the Hamiltonian decomposes into two sectors where the  Cooper-pair background field is that of a QPS with winding number $\nu_{c}\pm\nu_{s}$ respectively. If $\nu_{c}$ and $\nu_{s}$ are both half-integer, then the field in one of the sectors has odd winding and the other has even winding. Therefore there is a single zero mode on such a combined QPS. This zero mode is protected by the particle-hole symmetry operator $\mathcal{P}=\tau^z\rho^x\mathfrak{K}$.

\subsection{Bound states on merons}

At finite easy-plane anisotropy the relevant topological excitations that determine the properties of the spin sector of the model [Eq. (3)] are merons. As outlined in the main text a meron is essentially a vortex with a finite core in which the vector (otherwise confined to lie in the plane) escapes the plane and points in a direction perpendicular to the plane. This endows the meron with an additional degree of freedom: the orientation of the core. In other words, the meron carries two distinct topological charges, the vortex winding number $\nu_s$, and the skyrmion charge $Q$. The latter can only take the values $\pm \tfrac{1}{2}$.

Our goal is to find a physically more transparent derivation of the $\theta$-term in the action. Since in the this case the configuration of the superconducting phase is globally topologically trivial, it can be set to $\newtheta=0$, and we can parametrize the meron texture $\bm{n} = (\cos(\newtheta_s) \sin(\beta), \sin(\newtheta_s)\sin(\beta),\cos(\beta))$ by the in-plane angle $\newtheta_{s}$ and the out-of-plane angle $\beta$. We choose the meron to wind around the origin in the $xy$-plane, i.e. we set $\newtheta_s(r,\phi)=\nu_s\phi$. The lifting out of the plane is then described by $\beta$, which has to take the value $\pi/2$ far from the core, and $\pi(1 + 2Q)/2$ in the core, depending on the skyrmion charge $Q$. The nonzero components of the gap function are now given by $\Delta_{x}'=\abs{\Delta}k_{\mathrm{F}}\sin(\beta(r))\cos(\nu_s\phi)$, $\Delta_{y}'=\abs{\Delta}k_{\mathrm{F}}\sin(\beta(r))\sin(\nu_s\phi)$, and $\Delta_{z}'=\abs{\Delta}k_{\mathrm{F}}\cos(\beta(r))$. This yields the effective Hamiltonian
\begin{eqnarray}
H_{\mathrm{eff}}& =& -i\alpha^{r}v_{\mathrm{F}}\partial_{r} -i\frac{\alpha^{\phi}}{r}\partial_{\phi} - \abs{\Delta}k_{\mathrm{F}}\sin(\beta(r))(\cos(\nu_s\phi)\sigma^{x}\nonumber \\& &+\sin(\nu_s\phi)\sigma^{y})\tau^{y}\rho^{x}-\abs{\Delta}k_{\mathrm{F}}\cos(\beta(r))\sigma^{z}\tau^{y}\rho^{x}. \label{eq:HMerons}
\end{eqnarray}
If the anisotropy is strong, \emph{i.e.} the core size $L_{\rm an}\sim\sqrt{v_{s}/\lambda}$ is small, then it is natural to decompose this Hamiltonian into a unperturbed $H_{\mathrm{eff}}^{0}$ that contains the vortex winding of the meron, and a perturbation $V$ due to the core region
\begin{eqnarray}
H_{\mathrm{eff}}^{0}& =& -i\alpha^{r}v_{\mathrm{F}}\partial_{r} -i\frac{\alpha^{\phi}}{r}\partial_{\phi} - \abs{\Delta}k_{\mathrm{F}}\sin(\beta(r))(\cos(\nu_s\phi)\sigma^{x} \nonumber \\ & & +\sin(\nu_s\phi)\sigma^{y})\tau^{y}\rho^{x}, \\
V &=&-\abs{\Delta}k_{\mathrm{F}}\cos(\beta(r))\sigma^{z}\tau^{y}\rho^{x}. 
\end{eqnarray}
The unperturbed Hamiltonian $H_{\mathrm{eff}}^{0}$ can again be decomposed into a direct sum of Fu-Kane Hamiltonians. Indeed it possesses a time-reversal symmetry that squares to $-1$, given by $\mathcal{T}=\sigma^{x}\rho^{y}\mathfrak{K}$. The perturbation explicitly breaks this symmetry. Therefore, the unperturbed problem possesses two zero-modes for odd winding number, and the splitting effected by the perturbation must be symmetric (due to the still intact particle-hole symmetry).

For $\nu_s=+1$ the two (unperturbed) zero mode solutions read
\begin{eqnarray}
\Psi_{1}^{T}(r) &=& g_{s}(r)\frac{1}{\sqrt{2}}\begin{pmatrix}0&0& 0&0& 1&0& 0& 1\end{pmatrix}, \\
\Psi_{2}^{T}(r) &=& g_{s}(r)\frac{1}{\sqrt{2}}\begin{pmatrix}0&1& 1&0& 0&0& 0& 0\end{pmatrix},
\end{eqnarray}
where $g_{s}(r)= N\exp\left(-\int_{0}^{r}\mathrm{d}r'\frac{\abs{\Delta}k_{\mathrm{F}}}{v_{\mathrm{F}}}\sin\beta(r)\right)$ is a real function that is exponentially localized near $r=0$. In the space of the zero modes the perturbation takes the form
\begin{align}
V_{zm} &= \pi \abs{\Delta}k_{\mathrm{F}}\int_{0}^{\infty}\mathrm{d}rrg_s(r)^{2}\cos\beta(r)\begin{pmatrix}
0 & -i \\ i & 0
\end{pmatrix} \notag \\
& \sim \begin{cases} Q (\abs{\Delta} \kf) (L_{\rm{an}}/\xi)^2, & L_{\rm{an}} \ll \xi,\\ Q (\abs{\Delta}\kf), & L_{\rm{an}} \gg \xi.\end{cases}
\end{align}
In first order perturbation theory the two zero modes are split symmetrically around zero - the perturbative calculation is justified as long as the splitting is small compared to higher Caroli-de Gennes-Matricon states, which results in the condition $L_{\rm{an}} \ll \sqrt{  \abs{\Delta}/v_{\rm F}}$. 
In this limit one may neglect the coupling to higher energy states which implies that the kernel is block diagonal with one block consisting of the zero mode subspace and the other of the finite energy states (the restriction of $\mathcal{K}$ to this subspace is denoted $\mathcal{K}'$). Then the Pfaffian factorizes
\begin{equation}
\operatorname{Pf}\begin{pmatrix} C \mathcal K' & 0\\ 0 & V_{zm}\end{pmatrix} = \operatorname{Pf}\left( C \mathcal K'\right)\operatorname{Pf}\left(V_{zm}\right).
\end{equation} 
The Pfaffian of $V_{zm}$, a $2\times 2$-matrix, is given by the upper right off-diagonal element. Thus $\operatorname{Pf}V_{zm}=-iaQ$, where $a$ is a real and positive number. It can be seen that Pfaffians of configurations with opposite skyrmion charge $Q$ have opposite sign. Hence the contributions of merons and antimerons to the effective fugacity cancel and can be written in the form
\begin{equation}
    \operatorname{Pf}\left( C \mathcal K\right) \sim e^{-i\pi Q}.
\end{equation}
The relative phase produced by the fermions of the complex fugacity of configurations with different skyrmion charge $Q$ is therefore identical to the phase produced by a $\theta$-term with topological angle $\theta=\pi$.

\subsection{Numerical Results}

We also verify the existence of zero modes by numerical computation. To this end, the fermionic kernel is discretized on a lattice. Then the determinant (which is the square of the Pfaffian) is computed with the background field configuration of interest inserted. Due to the finite size of the lattice and due to the finite numerical accuracy it is not possible to show that the determinant in a certain background field vanishes exactly. However, we demonstrate two effects that strongly suggest the presence of zero modes: (i) The numerical value of the determinant in an appropriate topologically non-trivial background field is multiple orders of magnitude smaller compared to the determinant in a homogeneous configuration. (ii) In the background of a QPS-anti-QPS pair the determinant scales exponentially with the separation $\ell$ of the defects $\det(C\mathcal{K})\sim e^{-\ell/\xi}$. This is evidence for the linear confinement of QPS-anti-QPS pairs, in distinction to the usual logarithmic interaction.

\begin{figure}
    \centering
    \includegraphics[width = .45\textwidth]{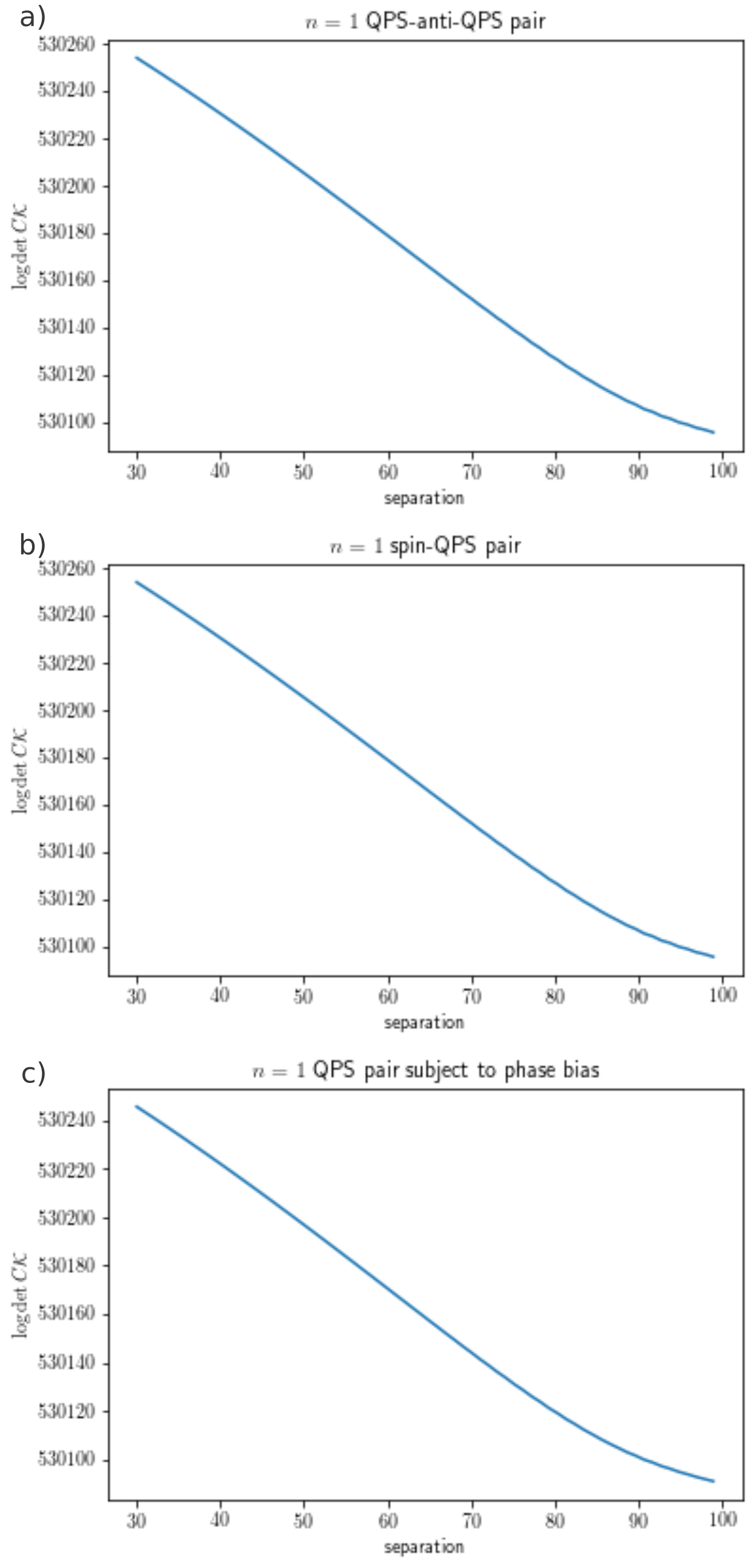}
    \caption{Numerical results for the fermion determinant. The logarithm of the determinant of the fermionic kernel is plotted against the separation between two topological defects. As expected from the presence of zero modes, the logarithm diminishes linearly with the separation. When a defect comes too close to the edge of the system, it hybridizes with the zero mode on the edge, and the linear behavior is altered. (a) QPS-anti-QPS pair with winding number $n=1$. (b) spin QPS-anti-QPS pair with winding number $n=1$. (c) QPS-anti-QPS pair with winding number $n=1$ subject to a phase bias of $2\pi$ along the spatial direction.}
    \label{fig:numerics}
\end{figure}

Concretely, we consider the kernel $C\mathcal{K}$ deep in the topological phase, Eq.~\eqref{eq:kernel_lin}. It is placed on a space-time lattice of $n_\tau=100$ sites along the imaginary-time direction and $n_x=300$ sites along the spatial direction. A QPS-anti-QPS pair with winding number $n=1$ is used as a background configuration and the separation between the defects is varied. 
For an exemplary separation of $50$ lattice sites the determinant in the non-trivial background is about $10^{58}$-times smaller compared to the determinant in the homogeneous background. Furthermore, the expected linear scaling of the logarithm of the determinant is observed, until the zero modes begin to hybridize with the edge modes. The same scaling behavior is observed for a pair of spin-QPS and a QPS pair subject to phase bias, Figs~\ref{fig:numerics} (b),(c).

\section{Computation of fermion correlators}

As outlined in the main text the presence of Fermi zero modes in an instanton background leads to algebraic decay of fermion correlators. This result can also be understood as analogous to the problem of chiral symmetry breaking in gauge theories: The fermion parity symmetry (which is separately conserved in the two spin sectors) does not allow gapless excitations with charge $e$ naively. But taking into account the instanton sectors renders this symmetry anomalous. In the following sections one- and two-fermion correlators are computed by considering contributions from zero modes in topologically nontrivial background fields. For simplicity we assume that $v_c=v_s$ in the following sections.

\subsection{Inclusion of Gauge Fields}

Before we proceed we briefly comment on the role of fluctuating electromagnetic 
fields in charged systems.
There are two effects, both of which only affect the charge field $\vartheta$, but leave the theory in spin-space 
unchanged: First, the logarithmic interaction of charge QPSs is formally screened beyond a space-time scale $L$~\cite{Duan1995} which is, however, exponentially large in the square of ratio of the London penetration depth and some microscopic scale.~\cite{ZaikinReply,konig2021resistance}
Second, the phase mode hybridizes with the 1D plasmon mode and decays faster than powerlaw because of a logarithmic momentum dependence of the stiffness~\cite{MooijSchoen1985}.
However, any metallic gate, located at a distance $d_s$ will screen the long-range interaction. While formally both phenomena would destroy algebraic superconductivity, it persists in the important and experimentally relevant regime  $d_s \ll x \ll L $.

\subsection{Single-fermion correlator}

The fermion correlator is given by a functional integral over all possible configurations of the fermions and the fluctuating order parameter weighted by the action
\begin{eqnarray}
 & &\langle\psi_{\sigma}(r)\psi_{\sigma'}^{\dagger}(0)\rangle \nonumber\\&=&  \sum_N\int \mathcal D\newtheta_N \mathcal D \bm{n}_N \mathcal D \psi \psi_{\sigma}(r)\psi_{\sigma'}^{\dagger}(0) e^{- S[\psi,\newtheta, \bm{n}]}. 
\end{eqnarray}
A given configuration of the order parameter can be divided into a singular vortex part and a regular fluctuation part $\newtheta=\newtheta^{v}+\newtheta^{sw},\,\newtheta_s=\newtheta_s^{v}+\newtheta_s^{sw},\beta=\beta^{v}+\beta^{sw}$. Let us focus on half QPS configurations, since they will generate the most relevant contribution to the correlator. We first perform a gauge transformation in order to separate the contribution of the fermions in the QPS background from the fluctuations. Introduce fermion operators $\chi_{\kappa}$ through
\begin{equation}
    \psi_{\sigma}(r) = e^{i\frac{\newtheta^{sw}(r)}{2}}U_{\sigma\kappa}(\bm{n}(r))\chi_{\kappa}(r),
\end{equation}
where $U_{\sigma\kappa}(\bm{n})$ denotes the elements of an $SU(2)$ matrix that satisfies
\begin{equation}
U^{\dagger}\bm{n}(\newtheta_s^{v}+\newtheta_s^{sw},\beta^{v}+\beta^{sw})\cdot\bm{\sigma} U = \bm{n}(\newtheta_s^{v},\beta^{v})\cdot\bm{\sigma},
\end{equation}
such that it transforms to a frame in which the fermions are subject only to the topologically nontrivial configuration. The matrix $U$ can be given explicitly as
\begin{widetext}
\begin{equation}
 U = e^{-i\frac{\newtheta_s^{sw}}{2}\sigma^{z}}e^{-i\frac{\beta^{sw}}{2}(\sigma^{y}\cos\newtheta_s^{v}+\sigma^{x}\sin\newtheta_s^{v})}  = \begin{pmatrix}
 e^{i\frac{\newtheta_s^{sw}}{2}}\cos\left(\frac{\beta^{sw}}{2}\right) & -e^{i\left(\frac{\newtheta_s^{sw}}{2}+\newtheta_s^{v}\right)}\sin\left(\frac{\beta^{sw}}{2}\right)\\ e^{-i\left(\frac{\newtheta_s^{sw}}{2}+\newtheta_s^{v}\right)}\sin\left(\frac{\beta^{sw}}{2}\right) & e^{-i\frac{\newtheta_s^{sw}}{2}}\cos\left(\frac{\beta^{sw}}{2}\right)
 \end{pmatrix} \label{eq:su2_matrix}
\end{equation} 
However, it cannot be expressed solely through the fluctuating angles, but it depends explicitly on the underlying vortex configuration through $\newtheta_s^{v}$. This has to be taken into account when the averages are performed. We therefore find
\begin{equation}
   \langle\psi_{\sigma}(r)\psi_{\sigma'}^{\dagger}(0)\rangle = \langle e^{i(\newtheta^{sw}(r)-\newtheta^{sw}(0))/2}\rangle  \sum_{n,\bm{d}_{i}}\frac{1}{(n!)^{2}}\int\mathrm{d}X_{1}\dots\mathrm{d}X_{2n}\mathcal{D}\chi \langle U_{\sigma\kappa}(\bm{n}(r))U^{\dagger}_{\kappa'\sigma'}(\bm{n}(0))\rangle \chi_{\kappa}(r)\chi_{\kappa'}^{\dagger}(0) e^{- S[\chi,\newtheta^{v}, \bm{n}^{v}]}, \label{eq:fermion_correlator}
\end{equation}
\end{widetext}
where the sum goes over $n$ QPS-anti-QPS pairs at the locations $X_{1},\dots,X_{2n}$ with orientation $\bm{d}_{i}$. Hence it can be seen that the correlator factorizes into a sum of fermion correlators in QPS-anti-QPS background fields and fluctuation contributions from $\newtheta$ and $\bm{n}$. (Of course even though the $2\pi$ combined QPSs are confined they still have to be summed over in the functional integral.)  

Let us now turn to the computation of the correlator of $\chi$-fermions. As outlined above it involves a sum over singular (QPS) background field configurations. We make the following approximation: we consider only the contribution of a single QPS-anti-QPS pair (of charge-spin combined QPSs) situated at distance $r$ (or more precisely, in a region of size $\xi$ around $r$) and at $0$. Then there is a localized (in space-time) zero mode with spin $\kappa$ at $r$ and a localized zero mode with spin $\kappa'$ at $0$. Taking into account configurations with higher numbers of QPS-anti-QPS pairs, these zero modes will eventually broaden into a band~\cite{diakonov_instantons_2003,schafer_instantons_1998} with the Fermi energy at half-filling, and the near-zero modes will become extended states. Whether the approximation above is justified depends on which order-parameter configurations dominate the path integral, dilute QPS-anti-QPS pair configurations, or dense ones. In any case, since the contributions of the extended zero modes to the correlator will also decay algebraically, they do not change the general conclusion of the presence of gapless fermion excitations.

Going back to the single QPS-anti-QPS pair, the expectation value $\langle\chi_{\kappa}(r)\chi_{\kappa'}^{\dagger}(0)\rangle$ in the background field is a constant independent of $r$ because of the integration over the (approximate, exact in the limit $r\rightarrow\infty$) Fermi zero modes. This can be seen by noting that the functional integral is defined as an integration over Grassmann numbers. Expanding the Nambu spinor fields $\Psi$ and its transposed $\Psi^{T}$ in some basis with Grassmann numbers $a_{n}$
\begin{align*}
\Psi(r) &= \sum_{n}\eta_{R,n}(r)a_{n}, \\
\Psi^{T}(r) &= \sum_{n}\eta_{L,n}^{T}(r)a_{n},
\end{align*}
where  the left and right basis functions can be chosen to be mutually orthonormal $\langle\eta_{L,n}\vert\eta_{R,m}\rangle=\delta_{n,m}$, the Majorana-Grassmann integral takes the form
\begin{equation}
\int\prod_{n}\mathrm{d}a_{n} \sum_{m,k}\eta_{R,m,1}(r)\eta_{L,k,1}^{T}(r)a_{m}a_{k} \prod_{l,s}\left(1-a_{l}\mathcal{K}_{ls}a_{s}\right),
\end{equation}
where $\mathcal{K}_{ls}$ denotes the matrix elements of the kernel in the introduced basis. From the preceding sections it is known that without the additional insertions the Grassmann integral vanishes, since it is equal to the Pfaffian $\operatorname{Pf}(\mathcal{K})$. From the point of view of Grassmann integration this means that there are Grassmann numbers that do not appear in the integrand. These are the zero modes. Hence the Grassmann integral becomes non-zero only if there is an insertion for every zero mode. From this it can be concluded that the expectation value above takes a finite value in the limit $r\rightarrow\infty$. The long-distance behavior now comes from the instanton action $\sim\exp\left(-\frac{K_{s}+K_{c}}{2}\log r\right)$ (this is just the interaction energy of the QPS and the AQPS). Of course combined QPSs with higher winding number $\nu$ will also contribute to the correlator, but their contribution will be subleading, because of the higher interaction energy that leads to the exponent $\nu^{2}(K_{s}+K_{c})/2$.

The fluctuation contributions of the superconducting phase are given by $\langle e^{i(\newtheta^{sw}(r)-\newtheta^{sw}(r'))/2}\rangle\sim r^{-1/8K_{c}}$. In a state without superconducting order this correlator decays exponentially. Hence the loss of phase coherence renders the fermions massive. Even though the zero modes are still present (locally) they do not influence the long-range properties anymore.

The fluctuation contribution of the $\bm{n}$-vector is encoded in the correlator $\langle U_{\sigma\kappa}(\bm{n}(r))U^{\dagger}_{\kappa'\sigma'}(\bm{n}(0))\rangle$. Using Eq.~\eqref{eq:fermion_correlator} we can write
\begin{widetext}
\begin{equation}
\langle\psi_{\sigma}(r)\psi_{\sigma'}^{\dagger}(0)\rangle = \langle e^{i(\newtheta^{sw}(r)-\newtheta^{sw}(0))/2}\rangle_{sw} \sum_{\kappa,\kappa'}\left\langle \int\mathcal{D}\chi \chi_{\kappa}(r)\chi_{\kappa'}^{\dagger}(0)\langle U_{\sigma\kappa}(\bm{n}(r))U^{\dagger}_{\kappa'\sigma'}(\bm{n}(0))\rangle_{sw} \right\rangle_{v},
\end{equation}
\end{widetext}
where $\langle\dots\rangle_{v(sw)}$ denotes the average over vortex (fluctuation) configurations. Taking $\sigma=\uparrow,\sigma'=\uparrow$ we find that three of the four terms in the sum depend on $\newtheta_s^{v}$. What happens to these terms upon averaging? The relevant contributions of the Grassmann integral come from close to the vortex cores. There the phase texture $\newtheta_s^{v}$ varies rapidly. Hence all terms involving $\newtheta_s^{v}$ will be suppressed. The remaining contribution is
\begin{eqnarray}
& &\langle \langle\chi_{\uparrow}(r)\chi_{\uparrow}^{\dagger}(0)\rangle\rangle_{v}\nonumber\\ & \times&\langle  e^{i\frac{\newtheta_s^{sw}(r)-\newtheta_s^{sw}(0))}{ 2}}\cos\frac{\beta^{sw}(r)}{2}\cos\frac{\beta^{sw}(0)}{2}\rangle_{sw}.
\end{eqnarray}
In the anisotropic case this is now easy to evaluate and we find
\begin{equation}
\langle\psi_{\sigma}(r)\psi_{\sigma'}^{\dagger}(0)\rangle \sim \delta_{\sigma\sigma'}r^{-1/8K_{c}-1/8K_{s}-K_{c}/2-K_{s}/2}.
\end{equation}
In the isotropic case $\newtheta_s^{sw}$ and $\beta^{sw}$ cannot be regarded as independent fields. Their dynamics must be described by a $SU_{1}(2)$ WZNW model. This can be effected by the replacement $K_s\rightarrow 1/2$. Hence
\begin{equation}
\langle\psi_{\sigma}(r)\psi_{\sigma'}^{\dagger}(0)\rangle \sim \delta_{\sigma\sigma'}r^{-1/8K_{c}-K_{c}/2-1/2}.
\end{equation} 

\subsection{Two-fermion correlators}

As we have seen above, the summation over combined QPSs (that host a single localized Fermi zero mode) leads to algebraic decay of single-fermion correlators. Analogously, configurations with two zero modes localized at the same point lead to algebraic decay of two-fermion correlators. This is an effect that does not originate from the presence of gapless single fermion excitations, which is why it persists in the vestigial phases. It has to be distinguished from the possible effect of extended single zero modes which appear for dense configurations of QPS-AQPS pairs. Essentially, if the single Fermi zero modes hybridize and form a band of finite width, then two fermion correlators are naturally gapless with a scaling dimension that is inherited from the single-fermion correlator. In the opposite limit of a very dilute QPS-AQPS pair configuration this is not implied. In this case the two-fermion correlator can only become gapless through order-parameter configurations with two zero modes localized at the same point.

Consider a $2\pi$ charge QPS. It possesses two localized zero modes of opposite spin, because the tunneling event changes the fermion parity in both the spin-up and -down sectors. This pair of zero modes forms a singlet, since it is unchanged by rotations in spin space. (Therefore the corresponding operator represents a type of order differing from the primary triplet order $\bm{\Delta}$.) The annihilation operator with the same quantum numbers (charge $2e$ and spin $0$) is given by $\psi_{\sigma}(r)\psi_{-\sigma}(r)$. Hence in the approximation that only configurations containing a single QPS-AQPS pair contribute, we obtain
\begin{widetext}
\begin{eqnarray}
	& &\langle (\psi_{\sigma}\psi_{-\sigma})(x)(\psi^{\dagger}_{-\sigma}\psi^{\dagger}_{\sigma})(0) \rangle \nonumber\\ &=& \langle e^{i(\newtheta^{sw}(r)-\newtheta^{sw}(0))}\rangle \times\langle \chi_{\sigma}\chi_{-\sigma}(r)\chi^{\dagger}_{-\sigma}\chi^{\dagger}_{\sigma}(0) \langle\cos^{2}\left(\frac{\beta^{sw}(r)}{2}\right)\cos^{2}\left(\frac{\beta^{sw}(0)}{2}\right)\rangle_{sw} \nonumber\\ & &- \chi_{\sigma}\chi_{-\sigma}(r)\chi^{\dagger}_{\sigma}\chi^{\dagger}_{-\sigma}(0) \langle\cos^{2}\left(\frac{\beta^{sw}(r)}{2}\right)\sin^{2}\left(\frac{\beta^{sw}(0)}{2}\right)\rangle_{sw} - \chi_{-\sigma}\chi_{\sigma}(r)\chi^{\dagger}_{-\sigma}\chi^{\dagger}_{\sigma}(0) \langle\sin^{2}\left(\frac{\beta^{sw}(r)}{2}\right)\cos^{2}\left(\frac{\beta^{sw}(0)}{2}\right)\rangle_{sw} \nonumber\\ & &+ \chi_{-\sigma}\chi_{\sigma}(r)\chi^{\dagger}_{\sigma}\chi^{\dagger}_{-\sigma}(0) \langle\sin^{2}\left(\frac{\beta^{sw}(r)}{2}\right)\sin^{2}\left(\frac{\beta^{sw}(0)}{2}\right)\rangle_{sw} \rangle_{v}.
\end{eqnarray}
\end{widetext}
Using the anticommutativity of the Grassmann numbers it turns out that the angular contributions cancel out. We find therefore
\begin{equation}
	\langle (\psi_{\sigma}\psi_{-\sigma})(r)(\psi^{\dagger}_{-\sigma}\psi^{\dagger}_{\sigma})(0) \rangle \sim r^{-1/2K_{c}-2K_{c}}.
\end{equation}
Because this correlator is independent of the variables in the spin sector, it remains gapless in the spin disordered vestigial charge-$4e$ phase.

We can also consider a $2\pi$ spin QPS. It hosts two zero modes of opposite spin, but now the phase is advanced in the opposite direction in both spin sectors, which implies that the fermion parity changes in the opposite direction. The corresponding operator thus has to be neutral and carry spin-$1$ (the spin QPS like $\bm{n}$ under spin rotations). Such an operator is given by $\psi_{\sigma}^{\dagger}\psi_{-\sigma}$. Note that the zero modes are only present in the limit of infinite anisotropy, otherwise they are gapped out. In this limit we find
\begin{eqnarray}
	\langle (\psi_{\sigma}^{\dagger}\psi_{-\sigma})(r)(\psi_{-\sigma}^{\dagger}\psi_{\sigma})(0)\rangle &\sim& \langle e^{-\sigma i (\newtheta_s^{sw}(r)-\newtheta_s^{sw}(0))} \rangle\nonumber\\ & \sim& r^{-1/2K_{s}-2K_{s}}.
\end{eqnarray}
It is independent of the charge variables. Hence at infinite anisotropy the charge disordered vestigial spin-$2$ phase possesses gapless spin-$1$ excitations.

\section{Non-Abelian bosonization}

In the following, a detailed derivation of the $\theta$-term using non-Abelian bosonization~\cite{witten_non-abelian_1984} is presented. Non-Abelian bosonization is a technique by which a $(1+1)$-dimensional system of fermions with a $O(N)\times O(N)$ (or $U(N)\times U(N)$) chiral symmetry can be rewritten as a fixed point of a bosonic theory, with the bosonic field being defined on the group manifold $O(N)$ ($U(N)$). The advantage of the approach, compared to Abelian bosonization, lies in the manifest preservation of the complete chiral symmetry. The action governing the group-valued field $g$ is the WZNW action of level $k=1$
\begin{equation}
S_W[g] = \frac{1}{4\gamma^{2}}\int\mathrm{d}^{2}x\operatorname{tr}\left[ \partial_{\mu}g \partial^{\mu}g^{-1} \right] + k\Gamma[g],
\end{equation}
where the Wess-Zumino term is defined extending the domain of $g$ to a hemisphere of the three-dimensional unit sphere with the original domain $S^{2}$ as its boundary
\begin{equation}
\Gamma[g] = -\frac{i}{24\pi} \int_{S^{3}_{\pm}}\mathrm{d}^{3}y \varepsilon^{\mu\nu\rho} \operatorname{tr}\left[ g^{-1} \frac{\partial g}{\partial y^{\mu}} g^{-1} \frac{\partial g}{\partial y^{\nu}} g^{-1} \frac{\partial g}{\partial y^{\rho}} \right]. \label{eq:WessZumino}
\end{equation}
By virtue of Stokes' theorem the Wess-Zumino term is independent of the precise choice of the extension.

In the case discussed here, the Lagrangian of fermions with spin in the background of a $p$-wave pairing field deep in the topological phase is expressed as a Lagrangian of Majorana fermions with a $O(4)\times O(4)$ chiral symmetry that is manifestly broken by the pairing terms that act as space-time dependent Majorana masses. In this form the free theory can be represented by a WZNW action of level $k=1$, and the pairing terms become mass terms for the group valued bosonic field $g$ that effectively pin $g$ to a fixed value. Under adiabatic variation of the gap function $g$ traces out a submanifold $\simeq (S^{1}\times S^{2})/\mathbb{Z}_{2}$ of $O(4)$. We demonstrate that the Wess-Zumino term~\eqref{eq:WessZumino} measures the skyrmion charge $Q$ of the pairing field configuration, $\Gamma[g(\newtheta,\bm{n})]=i\pi Q$, and is hence responsible for the appearance a $\theta$-term with $\theta=\pi$ in the effective theory describing the superconductor.

\subsection{Majorana representation of the free fermion action}

We begin by representing the free fermion action as a theory of massless Majorana fermions. Define the right- and left-moving fermion fields $R$ and $L$, and the right- and left-moving spinors as
\begin{equation}
\phi_{\rm{R}} = \begin{pmatrix}R\\ i\sigma^{y}L^{*}\end{pmatrix}, \quad \phi_{\rm{L}} = \begin{pmatrix}L\\ i\sigma^{y}R^{*}\end{pmatrix}.
\end{equation} 
After expansion around the Fermi points the BdG Hamiltonian [Eq. (1)] without pairing terms takes the form
\begin{equation}
\mathcal{H} = \frac{1}{2}\int\mathrm{d}x\left( \phi_{\rm{R}}^{\dagger}H_{\rm{R}}\phi_{\rm{R}} + \phi_{\rm{L}}^{\dagger}H_{\rm{L}}\phi_{\rm{L}} \right),
\end{equation}
with $H_{\rm{R}} = v_{\rm{F}}(-i\partial_{x})\tau^{z}$ and $H_{\rm{L}}=-H_{\rm{R}}$. The spinors are connected by the reality conditions $\phi_{\rm{R}}^{\dagger}=\phi_{\rm{L}}^{T}C$ and $\phi_{\rm{L}}^{\dagger}=\phi_{\rm{R}}^{T}C$, where $C=\sigma^{y}\tau^{y}$. Hence the Hamiltonian has the same number of degrees of freedom as that of $8$ Majorana fermions, but expressed through $\phi_{\rm{R}}$ and $\phi_{\rm{L}}$ it does not yet have the appropriate form. The desired Nambu spinor should satisfy the Majorana condition $\Phi^{\dagger}=\Phi^{T}$. This condition is satisfied by the spinor $\chi^{T} = (\chi_{1\rm{R}}^{T},\chi_{2\rm{R}}^{T},\chi_{1\rm{L}}^{T},\chi_{2\rm{L}}^{T})$ defined through
\begin{subequations}
\label{eq:majorana}
\begin{eqnarray}
R_{\sigma} &=& \frac{1}{\sqrt{2}}\left( \chi_{\sigma1\rm{R}}+i\chi_{\sigma2\rm{R}},\right) \\
\sigma^{y}_{\sigma\sigma'}L_{\sigma'} &=& -\frac{1}{\sqrt{2}}\left(\chi_{\sigma1\rm{L}}-i\chi_{\sigma2\rm{L}}\right).
\end{eqnarray}
\end{subequations}

Under this transformation the kinetic part of the Hamiltonian becomes $H_{\rm{kin}}=v_{\rm{F}}i\partial_{x}\rho^{z}$. 

It remains to express the pairing terms in this representation. Before the transformation they are given by $H_{\rm{p,R}}=\Delta_{j}'\sigma^{j}\tau^{y}+\Delta_{j}''\sigma^{j}\tau^{x}$ and $H_{\rm{p,L}}=-H_{\rm{p,R}}$. The result after the transformation is
\begin{eqnarray}
H_{\rm{p}} &= &\Delta_{x}'\sigma^{x}\tau^{y}\rho^{x} + \Delta_{x}''\sigma^{x}\tau^{0}\rho^{y} - \Delta_{y}'\sigma^{y}\tau^{0}\rho^{x} - \Delta_{y}''\sigma^{y}\tau^{y}\rho^{y} \nonumber\\ & &+ \Delta_{z}'\sigma^{z}\tau^{y}\rho^{x} + \Delta_{z}''\sigma^{z}\tau^{0}\rho^{y}. \label{eq:pair_ham}
\end{eqnarray}
It can be seen that this BdG Hamiltonian only contains off-diagonal $\rho$-matrices. Hence it only couples left- and right-moving fermions. Therefore it effect is to generate a mass term linear in $g$ upon bosonization.

\subsection{Bosonization}

Now, the group valued field $g$ is introduced. The correspondences between the currents are given by
\begin{subequations}
\label{eq:bosonization}
\begin{eqnarray}
\chi_{\rm{L}}^{i}\chi_{\rm{L}}^{j} &\longleftrightarrow &-\frac{1}{2\pi}\left(g^{-1}\partial_{+}g\right)^{ij}, \\
\chi_{\rm{R}}^{i}\chi_{\rm{R}}^{j} &\longleftrightarrow &\frac{1}{2\pi}\left((\partial_{-}g)g^{-1}\right)^{ij},\\
\chi_{\rm{R}}^{i}\chi_{\rm{L}}^{j} &\longleftrightarrow &iMg^{ij},
\end{eqnarray}
\end{subequations}
where $M$ is a quantity with dimensions of mass that depends on the regularization procedure.

These formulas can now be used to represent the pairing terms through $g$. Parametrize the order parameter as above. Consider e.g. the first contribution to Eq.~\eqref{eq:pair_ham}. It holds
\begin{eqnarray}
\chi^{T}\tau^{y}\sigma^{x}\rho^{x}\chi &=& \chi_{\rm{R}}^{T}\tau^{y}\sigma^{x}\chi_{\rm{L}} + \chi_{\rm{L}}^{T}\tau^{y}\sigma^{x}\chi_{\rm{R}} \nonumber\\ &=& iM\left( g^{ij}(\tau^{y}\sigma^{x})_{ij} - g^{ji}(\tau^{y}\sigma^{x})_{ij}\right) \nonumber\\ &=& -iM \mathrm{tr}\left[g \left(\tau^{y}\sigma^{x}-(\tau^{y}\sigma^{x})^{T}\right)\right] \nonumber\\ &=& -2iM \mathrm{tr}\left[ g\tau^{y}\sigma^{x}\right].
\end{eqnarray}
Proceeding analogously for the remaining terms yields
\begin{equation}
S_{\rm{p}}[g,\newtheta,\bm{n}] = -\abs{\Delta}k_{\mathrm{F}}M \operatorname{tr}\left[ g O_{U}^{T} \right], \label{eq:mass_term}
\end{equation}
where $O_{U}$ defines an embedding of the order parameter manifold into the target space of $g$
\begin{eqnarray}
(S^{1}\times S^{2})/\mathbb{Z}_{2}\rightarrow O(4): \,\, (\newtheta,\bm{n})\longmapsto O_{U}, \nonumber\\
O_{U} = -i\cos\newtheta \left(n_{x}\tau^{y}\sigma^{x} - n_{y}\tau^{0}\sigma^{y} + n_{z}\tau^{y}\sigma^{z}\right) \nonumber\\ - \sin\newtheta \left(n_{x}\tau^{0}\sigma^{x} - n_{y}\tau^{y}\sigma^{y} + n_{z}\tau^{0}\sigma^{z}\right). \label{eq:embedding}
\end{eqnarray}
It can be checked that $O_{U}$ is indeed orthogonal. Moreover its determinant is equal to one $\det(O_{U})=1$ which means that the mapping is into the $SO(4)$ component of $O(4)$.

\subsection{Mean-Field Solution}

For fixed background fields $\bm{n},\newtheta$ the pairing term is minimized by the configuration $g=O_{U}$. This can be seen by virtue of the following argument: The matrix $-gO_{U}^{T}$ is an element of $O(4)$. Its trace is given by the sum of its eigenvalues. Because of the orthogonality the possible eigenvalues lie on the complex unit circle, and, if they are not real, come in pairs of complex conjugates. Hence the minimal value of the trace is $-4$ which corresponds to $4$ eigenvalues of $-1$. Indeed a matrix with $4$ eigenvalues of $-1$ is equal to the negative of the identity matrix and hence also unique. 

The mean-field solution can be used as a starting point to find the effective action. As discussed above it parametrizes the Goldstone manifold. Fluctuations around this manifold are massive. Define $h=gO_{U}^{T}$ and use the Wiegmann-Polyakov formula~\cite{gogolin_bosonization_1998}
\begin{eqnarray}
	S_W[hO_{U}] &=& S_W[h] + S_W[O_{U}] \nonumber\\& &+ \frac{1}{2\pi} \int\mathrm{d}^{2}x \mathrm{tr} \left( h^{-1}\bar{\partial}h\, O_{U}\partial O_{U}^{T} \right), \label{eq:wiegmann}
\end{eqnarray} 
where $\bar{\partial}=\frac{1}{2}(\partial_{x}+i\partial_{\tau})$ and $\partial=\frac{1}{2}(\partial_{x}-i\partial_{\tau})$. $S_W[O_{U}]$ constitutes the leading order contribution to the effective action. The kinetic part gives
\begin{equation}
S_{\rm{kin}} = \frac{1}{\gamma^{2}}\int\mathrm{d}^{2}x \left( \left(\partial_{\mu}\newtheta\right)^{2} + \left(\partial_{\mu}\bm{n}\right)^{2} \right),
\end{equation}
which is simply the action of the NL$\sigma$M corresponding to the order parameter manifold. This constitutes a renormalization of the NL$\sigma$M due to the contribution of the fermions (the starting point is already a NL$\sigma$M with stiffnesses $K_{c,s}^{0}$). Its magnitude can be estimated by using the fixed point value of the $O(4)\times O(4)$ symmetric model as a bare value. Then $\gamma^{2}=4\pi$ and therefore the renormalization of the stiffness is $K_{\mathrm{f}}=1/2$.

Let us now determine the effect of the Wess-Zumino (WZ) term. In order to substitute the mean-field configuration into the WZ term it is necessary to extend it to the three-sphere $S^{3}$. Following~\cite{altland_condensed_2010} parametrize $S^{3}$ by the coordinates on $S^{2},\,(x_{1},x_{2})$ and an additional polar angle $\xi\in[0,\pi]$. The equator corresponds to $\xi=\pi/2$. Hence the extension $\tilde{g}$ has to satisfy $\tilde{g}\left(x_{1},x_{2},\frac{\pi}{2}\right)=g( x_{1},x_{2})$ and the limit $\tilde{g}(\xi\rightarrow 0)$ has to exist.

Since the only field that can obtain a nontrivial $\theta$-term from the WZNW action is the vector $\bm{n}$ it is sufficient to set the superconducting phase $\theta$ to a constant value. For simplicity $\theta=\pi/2$ is chosen. Then the semi-classical configuration that is to be extended reads
\begin{equation}
g = -\tau^{y}\bm{n}\cdot\bm{\Sigma},\quad \bm{\Sigma}\equiv\begin{pmatrix}\tau^{y}\sigma^{x}&-\sigma^{y}&\tau^{y}\sigma^{z}\end{pmatrix},
\end{equation}
where the vector of matrices $\bm{\Sigma}$ has been introduced. It satisfies the algebra
\begin{equation}
\Sigma_{i}\Sigma_{j} = \delta_{ij}1-i\epsilon_{ijk}\Sigma_{k}, \label{eq:algebra}
\end{equation}
which is analogous to the algebra of Pauli matrices up to a sign. Now, a concrete example of an extension is given by
\begin{equation}
\tilde{g} = i\tau^{y}\exp\left(i\xi \bm{n}\cdot\bm{\Sigma}\right). \label{eq:extension}
\end{equation}
This can be seen by noting that
\begin{equation}
\tilde{g} = i\tau^{y}\left( \cos\xi + i\bm{n}\cdot\bm{\Sigma}\sin\xi \right) = i\tau^{y}\cos\xi + g\sin\xi.
\end{equation}
This coincides with $g$ for $\xi=\pi/2$ and is an element of $O(4)$ on the whole three-sphere.

The expression~\eqref{eq:extension} can now be inserted into the WZ term. Defining $\bm{\omega}_{i}\equiv \bm{n}\times\partial_{i}\bm{n}$ it can be expressed as
\begin{eqnarray}
 \Gamma[g=O_{U}] &=& -\frac{6i}{24\pi} \int_{0}^{\frac{\pi}{2}}\mathrm{d}\xi \left(-i\sin^{2}\xi\right) \int\mathrm{d}^{2}x \nonumber\\&  &\times\mathrm{tr}[ \bm{n}\cdot\bm{\Sigma} ( \cos\left(\xi\right) \partial_{1}\bm{n} + \sin\left(\xi\right) \bm{\omega}_{1} )\cdot\bm{\Sigma} \nonumber\\ &  &\times( \cos\left(\xi\right) \partial_{2}\bm{n} + \sin\left(\xi\right) \bm{\omega}_{2} )\cdot\bm{\Sigma} ].
\end{eqnarray}
By repeatedly applying Eq.~\eqref{eq:algebra} and using that the $\Sigma_{i}$ are traceless this can be brought into the form
\begin{eqnarray}
&-&\frac{6i}{24\pi} \int_{0}^{\frac{\pi}{2}}\mathrm{d}\xi (-4)\sin^{2}\xi \int\mathrm{d}^{2}x \cos^{2}\xi\,\bm{n}\cdot\left(\partial_{1}\bm{n}\times\partial_{2}\bm{n}\right) \nonumber\\ &+& \sin^{2}\xi\, \bm{n}\cdot\left(\bm{\omega}_{1}\times\bm{\omega}_{2}\right) \nonumber\\&+& \sin\xi\cos\xi\,\bm{n}\cdot\left( \bm{\omega}_{1}\times\partial_{2}\bm{n} + \partial_{1}\bm{n}\times\bm{\omega}_{2} \right).
\end{eqnarray}
The third term vanishes
\begin{equation}
\bm{n}\cdot\left( \bm{\omega}_{1}\times\partial_{2}\bm{n}\right) = - \bm{\omega}_{1}\cdot\bm{\omega}_{2} = - \bm{n}\cdot\left(\partial_{1}\bm{n}\times\bm{\omega}_{2} \right),
\end{equation}
and the first two terms combine, $\bm{n}\cdot\left(\bm{\omega}_{1}\times\bm{\omega}_{2}\right)=\bm{n}\cdot\left(\partial_{1}\bm{n}\times\partial_{2}\bm{n}\right)$. The integral over $\xi$ evaluates to $\pi/4$ and therefore the result is
\begin{equation}
\Gamma[g=O_{U}] = \frac{6\pi i}{24\pi} \int\mathrm{d}^{2}x \bm{n}\cdot\left(\partial_{1}\bm{n}\times\partial_{2}\bm{n}\right). \label{eq:theta}
\end{equation}
But this can be identified as the $\theta$-term of a NL$\sigma$M defined on $S^{2}$~\cite{altland_condensed_2010}
\begin{equation}
S_{\theta} = \frac{i\theta}{4\pi}\int\mathrm{d}^{2}x \bm{n}\cdot\left(\partial_{1}\bm{n}\times\partial_{2}\bm{n}\right).
\end{equation}
Comparison with the result from the WZNW model shows that the $\theta$-angle takes the value $\pi$.

\subsection{BCS interactions}

In the following we demonstrate starting from a purely fermionic model with BCS interactions in the $p$-wave channel, that a state characterized by a superconducting order parameter of the form $\bm{\Delta}\sim e^{i\newtheta}\bm{n}$ is favored. We show that the family of mean-field states parametrized by Eq.~\eqref{eq:embedding} arises as the ground-state manifold of the interacting model.

Consider the local interaction operator
\begin{equation}
S_{\rm{int}} = -V \int\mathrm{d}^{2}x \left( \psi^{\dagger}\bm{\sigma}i\sigma^{y}\partial_{x}\psi^{*}\right) \left( \psi^{T}i\sigma^{y}\bm{\sigma}\partial_{x}\psi\right).              \label{eq:bcs}
\end{equation}
For positive interaction constant $V>0$ this interaction is attractive and drives a BCS instability in the $p$-wave channel. A mean-field decoupling yields the BdG Hamiltonian [Eq. (1)]. In order to bosonize this term, we expand the fermion operators close to the Fermi points and only consider momentum-conserving terms. At half-filling Umklapp scattering becomes relevant. This is neglected. There are two types of momentum-conserving scattering events: Two right- (left-) moving Fermi fields can scatter into to right- (left-) moving Fermi fields. After bosonization these terms will contain derivatives of the group-valued field $g$ and hence lead to a renormalization of the kinetic term. Hence they are not considered further. On the other hand, a right- and a left-mover can scatter into a right- and a left-mover. These processes lead to a potential term that reduces the gapless degrees of freedom. We obtain
\begin{eqnarray}
S_{\rm{int}} &=& Vk_{\mathrm{F}}^{2}\int\mathrm{d}^{2}x \left( R^{\dagger}\bm{\sigma}i\sigma^{y}L^{*} - L^{\dagger}\bm{\sigma}i\sigma^{y}R^{*} \right) \nonumber\\ &  &\times\left( -R^{T}i\sigma^{y}\bm{\sigma}L + L^{T}i\sigma^{y}\bm{\sigma}R \right).
\end{eqnarray}
Define $\tau^{\pm}\equiv\tau^{0}\pm \tau^{y}$. Transforming to the Majorana representation~\eqref{eq:majorana} yields
\begin{eqnarray}
S_{\rm{int}} &=& \frac{1}{4} Vk_{\mathrm{F}}^{2}\int\mathrm{d}^{2}x \left( i\chi_{R}^{T}\bm{\sigma}\tau^{-}\chi_{L} + i\chi_{L}^{T}\sigma^{y}\bm{\sigma}\sigma^{y}\tau^ {+} \right) \nonumber\\ &  &\times\left( i\chi_{R}^{T}\sigma^{y}\bm{\sigma}\sigma^{y}\tau^{+} + i\chi_{L}^{T}\bm{\sigma}\tau^{-}\chi_{R} \right).
\end{eqnarray}
In this form the interaction can be bosonized using Eqs~\eqref{eq:bosonization}. We define the effective interaction strength $\eta=Vk_{\mathrm{F}}^{2}M^{2}>0$ and find
\begin{equation}
S_{\mathrm{int}}[g] = Vk_{\mathrm{F}}^{2}M^{2} \int\mathrm{d}^{2}x \operatorname{tr}\left[g \sigma^{y}\bm{\sigma}\sigma^{y}\tau^{+}\right] \operatorname{tr} \left[ g\bm{\sigma}\tau^{-} \right]. \label{eq:bcs_bosonized}
\end{equation}

In order to show that the family of solutions parametrized by Eq.~\eqref{eq:embedding} indeed minimizes Eq.~\eqref{eq:bcs_bosonized}, we expand the action in fluctuations around this solution. Therefor we introduce the imaginary, antisymmetric, and traceless matrices $L_{ab} = -i(e_{ab}-e_{ba})$ for $a<b,\,a,b\in\{1,2,3,4\}$, where $e_{ab}$ has entry $1$ at the position $(ab)$ and zero everywhere else. These matrices form a basis of the Lie algebra $\mathfrak{so}(4)$. This allows to write $g=O_{U}\exp\left(\tfrac{i}{\sqrt{2}}\phi_{ab}L_{ab}\right)$, where $\phi_{ab}$ are fluctuating fields. Inserting this into the action~\eqref{eq:bcs_bosonized} and retaining only terms up to second order in the fields $\phi_{ab}$ yields
\begin{equation}
    S_{\mathrm{int}} = \eta\left(-16 + \frac{1}{2} C_{(ab)(cd)}(\newtheta,\bm{n})\phi_{ab}\phi_{cd}\right).
\end{equation}
The mass matrix $C_{(ab)(cd)}$ possesses three zero eigenvalues and three positive eigenvalues for all $(\newtheta,\bm{n})$. The three zero-eigenvalue modes correspond to translations along the manifold, while the three positive eigenvalues describe gapped fluctuations with mass $\sqrt{8\eta}$. Hence the solutions $g=O_{U}(\newtheta,\bm{n})$ indeed constitute the ground-state manifold of the interaction~\eqref{eq:bcs}, and it therefore favors an order parameter of the form $\bm{\Delta}\sim e^{i\newtheta}\bm{n}$.

\subsubsection{Hubbard-Stratonovich Transformation}

The fermion-fermion interaction operator Eq.~\eqref{eq:bcs} provides one specific starting point to obtain the two-fluid model Eq. (1). A Hubbard-Stratonovich decoupling yields the Hamiltonian [Eq. (1)] and integrating out amplitude fluctuations produces the NL$\sigma$M. Our results do not rely on  this concrete derivation, since the two-fluid model is the generic low-energy theory with the desired order-parameter symmetry. The NL$\sigma$M now encodes the soft fluctuations of the Goldstone manifold. If this manifold is unstable the model flows to strong coupling, indicating the absence of the ordered phase.
Let us make this explicit by performing the Hubbard-Stratonovich decoupling for this concrete example. We introduce the order parameter $\bm{\Delta}$ as a Hubbard-Stratonovich (HS) field and rewrite the exponential of the interaction operator as
\begin{widetext}
\begin{align}
    e^{-S_{\mathrm{int}}} \propto\int\mathcal{D}\bm{\Delta}^\dagger\mathcal{D}\bm{\Delta}  \exp\Bigg( -\int\mathrm{d}^2 x\bigg[ \frac{\bm{\Delta}^\dagger\cdot\bm{\Delta}}{V}  -\bm{\Delta}\cdot\left( \psi^{\dagger}\bm{\sigma}i\sigma^{y}\partial_{x}\psi^{*}\right) \bigg.\Bigg.  -
    \Bigg.\bigg.\bm{\Delta}^\dagger\cdot\left( \psi^{T}i\sigma^{y}\bm{\sigma}\partial_{x}\psi\right) \bigg]\Bigg).
\end{align}
\end{widetext}
The linear couplings between the HS field and the fermion bilinears generates the pairing terms of the BdG description. The NL$\sigma$M formulation is recovered upon fixing the amplitude of the HS field to a non-zero value. The bare stiffnesses of this model are equal to zero. The coupling to the fermions dynamically generates a finite renormalized stiffness. Performing a gradient expansion one finds $K_{c,s}^{\mathrm{ren}}=2F\left(\frac{\mu}{2m\abs{\Delta}^2}\right)$ with $F(x)\rightarrow 1$ for $x\rightarrow\infty$, $F(x<0)=0$, and $F(x)\sim x^{1/4}\log^{1/2}(1/x)$ for $x\rightarrow0^+$. Hence $K_{c,s}=2$ deep in the topological phase for this concrete model. On the mean-field level a self-consistent solution with a finite amplitude exists if the ground-state energy is lowered compared to the normal state. Consider the change in ground-state energy density
\begin{equation}
    \Delta\mathcal{E} = \int\frac{\mathrm{d}k}{2\pi}\left(\xi_k-\sqrt{\xi_k^2+\abs{\bm{\Delta}}^2k^2}\right) + \frac{\abs{\bm{\Delta}}^2}{V}.
\end{equation}
This quantity has to be negative for a nontrivial self-consistent solution to exist. Deep in the weak-paired phase this is certainly the case: expanding around the Fermi points and assuming a constant density of states yields the familiar BCS criterion.

\end{document}